\documentclass[a4paper,11pt, accepted=2022-08-31]{quantumarticle}

\pdfoutput=1
\usepackage[utf8]{inputenc}
\usepackage[english]{babel}
\usepackage[T1]{fontenc}
\usepackage{amsmath}
\usepackage{mathrsfs}
\usepackage{amsfonts}
\usepackage{amssymb}
\usepackage{amsmath}
\usepackage{graphicx}
\usepackage{color}
\usepackage{stackrel}
\usepackage{hyperref}
\usepackage{upgreek}

\newcommand{\Tr}{\mathrm{Tr}}

\newcommand{\ket}[1]{\vert #1 \rangle}

\newcommand{\braket}[2]{\langle #1 \vert #2 \rangle}

\newcommand{\be}{\begin{equation}}
\newcommand{\ee}{\end{equation}}
\newcommand{\bqa}{\begin{eqnarray}}
\newcommand{\eqa}{\end{eqnarray}}

\graphicspath{{./figures/}}

\begin{document}

\title{Policy Gradient Approach to Compilation of Variational Quantum Circuits}

\author{D. A. Herrera-Mart\'i}
\affiliation{Universit\'e Grenoble Alpes, CEA List, 38000 Grenoble, France}

\maketitle

\begin{abstract} 
We propose a method for finding approximate compilations of quantum unitary transformations, based on techniques from policy gradient reinforcement learning. The choice of a stochastic policy allows us to rephrase the optimization problem in terms of probability distributions, rather than variational gates. In this framework, the optimal configuration is found by optimizing over distribution parameters, rather than over free angles. We show numerically that this approach can be more competitive than gradient-free methods, for a comparable amount of resources, both for noiseless and noisy circuits. Another interesting feature of this approach to variational compilation is that it does not need a separate register and long-range interactions to estimate the end-point fidelity, which is an improvement over methods which rely on the Hilbert-Schmidt test. We expect these techniques to be relevant for training variational circuits in other contexts.
\end{abstract}

\section{Introduction}

The general problem of quantum compilation is to approximate any unitary transformation with a sequence of elements selected from a fixed universal set of quantum gates. The existence of an approximate sequence of quantum gates for a single qubit is guaranteed by the Solovay-Kitaev theorem \cite{NCbook}, which states that any single-qubit gate can be approximated with an overhead logarithmic in the original number of gates, i.e. polylogarithmic as $O(\log^c (1/\epsilon))$, where $\epsilon$ is the approximation accuracy and c is a constant lower-bounded by 1 \cite{harrow}.

Although the Solovay-Kitaev theorem proves that any computation can be efficiently approximated within an arbitrary tolerance, it does not tell us how to find the optimal sequence of gates. The standard algorithm uses an exhaustive search technique to build a library of gate sequences in its lowest level of recursion, and then builds on it recursively. In general, the longer the sequence of library gates (and their inverses), the better the approximation to the target unitary \cite{dawson}. 

Finding the optimal compilation of a quantum unitary is equivalent to finding the shortest path between two nodes (the geodesic) in a (hyperbolic) Cayley graph \cite{lin}. Hyperbolic graphs resemble tree graphs in the sense that for an overwhelming majority of node pairs,  there is only one path linking both nodes \cite{hyperbolic}. Therefore, the geometric intuition for the hardness of exact compilation is that, in a hyperbolic graph, looking for the shortest path involves evaluating an exponential number of nodes at each step. Indeed, performing an optimal compilation of a given quantum circuit is believed to be a hard problem under reasonable assumptions \cite{nielsen06science, lin}.

With the advent of sub-threshold quantum architectures \cite{preskill}, research on variational quantum algorithms has become central to the field of quantum computing, giving rise, among others, to alternative routes to universal quantum computation based on variational circuits \cite{lloyd, morales, kiani}. The existence of variational tasks which are believed to be intractable for classical computers \cite{fahri,arute,zhu,bravyi18,bravyi20,bauer} is an encouraging motivation for research in variational quantum algorithms. By considering a discretization of the rotation angles, which are the free variational parameters, it can be seen that the set of circuits that can be built using these gates has a Cayley graph which retains its hyperbolic character \cite{lin}, and therefore it is likely that finding good angle configurations will become too difficult for large unitaries.

Furthermore, several bottlenecks must be addressed before variational quantum algorithms on a NISQ processor can be properly trained at scale. These include the ability to efficiently evaluate cost functions which are made of non-commuting observables\cite{omalley, ralli}; and the design of new algorithms with no known efficient classical simulation \cite{hastings, bravyi20obstacles, bravyi20hybrid}, in which the optimizations of the variational parameters may suffer from the curse of dimensionality as the number of qubits increases.

A further impediment to VQA training is due to the fact that there are regions in parameter space, commonly known as \emph{barren plateaus}, where the cost gradient vanishes exponentially in the number of qubits. This behaviour of the cost function, which exponentially increases the resources required to train large-scale quantum neural networks, is a manifestation of the concentration of measure phenomenon and it has been demonstrated in a number of proposed architectures and classes of cost functions \cite{mcclean, cerezo}. Similarly, gradient-free optimizers are unable to cope with the barren plateau problem, as finite differences on which their iterative improvements are based, are exponentially suppressed in a barren plateau \cite{arrasmith}.

Several strategies have been suggested to mitigate the effect of barren plateaus during the training of variational quantum circuits. Some techniques aim at circumventing the probabilistic independence assumption which underlies the concentration results \cite{grant,volkoff}, while others try to train the circuit piecewise or using subsensembles of angles \cite{skolik, cerezo}. Although each of these methods has advantages and drawbacks, no fully satisfactory solution to this problem has yet been devised. In this work, we explore ideas from reinforcement learning, policy gradient methods in particular, to mitigate the effects of barren plateaus in the training of variational quantum algorithms of shallow depth (logarithmic in the number of qubits), and we apply them to the particular case of approximate compilation. The intuition behind this choice comes from the fact that in policy gradient algorithms, the cost function can be written as an expectation value of a parameterized analytic function. This means that an update rule can be defined which involves sampling potential configurations in the local neighbourhood of the current solution. The size of this neighbourhood can be tuned dynamically by changing the update stride (see Appendix A). We show that this approach is competitive and that it can outperform gradient-free methods in noiseless and noisy circuits at the onset of a barren plateau.

\section{Compilation of Variational Quantum Algorithms}

\begin{figure}[ht!]
  \includegraphics[width=\linewidth]{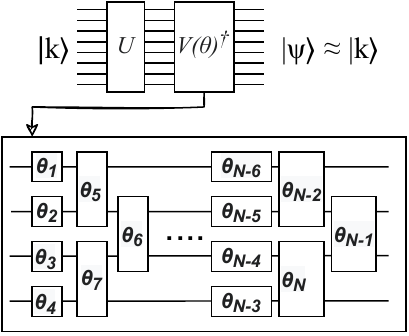}
  \caption{Circuit Setup. The goal is to retrieve a state with a maximal overlap with the initial state $\ket{k}$, for all initial states, randomly sampled from a fixed basis. The unknown unitary $U$ is followed by $V(\theta)^\dagger$, resulting in a state $\ket{\psi}$. At the end of the circuit, the measurement projects the state back onto the initial state with a probability that depends on the overlap, which constitutes the reward, i.e. $r^{(k)}_\theta = |\langle k|V(\theta)^\dagger U |k\rangle|^2$. The parameterized unitary is made of single qubit rotations of the form $R_Y(\theta_i) = \exp (- \textrm{i} Y \theta_i$) and two-qubit rotations are of the form $R_{ZZ} (\theta_i) = \exp (-\textrm{i} ZZ \theta_i$)}
  \label{fig:setup}
\end{figure}

The goal is to learn the action of an unknown unitary gate $U$ on an arbitrarily large set of initial states (see Fig. \ref{fig:setup}). The first assumption we make is that $U$ is at most of logarithmic depth, which is motivated by the fact that in NISQ architectures (in the absence of error correction) this is already a beneficial scaling, and by the fact that barren plateaus will arise for global cost functions, such as in Eq.~(\ref{eq:fidelity}) (\cite{cerezo}). Our second working hypothesis is that the interactions giving rise to the unknown unitary, i.e. the qubit connectivity graph, are known. We have assumed that they are nearest-neighbours, but more general interactions are straightforward, provided that the interaction graph is known to the compiler. In the absence of complete information regarding connectivity, an all-to-all circuit should be trained, which would entail a quadratic growth of the number of free parameters with the number of qubits. Having to optimize over extra spurious angles would reduce the interest in this method.

There are several approaches \cite{lloyd,kiani,sharma,khatri} to assess the performance of an approximate compilation method. We will adopt a variation of one metric introduced in Refs. \cite{sharma,khatri}, motivated by its experimental feasibility. The estimate: 

\be
\hat F(\theta) = \frac{1}{m}\sum^m_k |\langle k|V(\theta)^\dagger U |k\rangle|^2,
\label{eq:fidelity}
\ee
corresponds to the fidelity between the initial and the final states (after compilation), averaged over different initial states. To fully characterize $U$, a tomographically complete characterization demands $\mathcal{O}(4^{N_q})$ different initial states, for $N_q$ qubits. Variational quantum compilation of a full unitary matrix U by considering the action of U on a complete basis rapidly gets computationally demanding as the number of qubits grows. A simpler task is to learn to prepare only a particular column of the matrix U by considering the action of U on a fixed input state, or on a small subset of initial states. The Hilbert-Schmidt test in \cite{sharma,khatri} has the advantage that it estimates the gate similarity $\Tr[V(\theta)^\dagger U]$, at the cost of doubling the number of qubits and introducing highly non-local interactions. In this work, we will estimate the fidelity using only $m  \sim poly(N_q)$ initial states (see Appendix B), a fact which is denoted by the hat punctuation in the previous equation. 

Approximate compilation of random unitaries, phrased as a variational algorithm, will necessarily get stuck in barren plateaus \cite{holmes}. We will try to mitigate this problem by optimizing a stochastic version of the cost function based on Eq. (\ref{eq:fidelity}).

\section{Policy Gradient for Quantum Compilation}

\begin{figure}[ht!]
  \includegraphics[width=\linewidth]{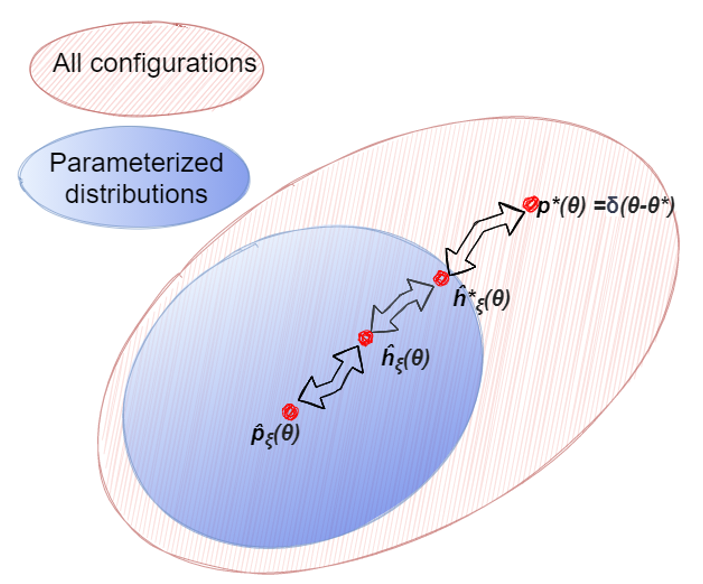}
  \caption{Expressivity of the model as a function of the chosen policy. A given set of probability distributions parameterized by some vector $\xi$ is contained in the larger set of all probability distributions. The statistical distance which separates the learned distribution from the ground truth has three components. The {\bf training error}, which underlies the optimization method, is the statistical distance between the current distribution $\hat p_\xi(\theta)$ and the empirically optimal hypothesis $\hat h_\xi(\theta)$. The distance between the empirically optimal hypothesis and the optimal hypothesis $\hat h^*_\xi(\theta)$ is given by the {\bf finiteness of sampling}. Finally, the distance between $\hat h^*_\xi(\theta)$ and the ground truth $p^*(\theta) = \delta (\theta - \theta^*)$ depends on the {\bf model expressivity} of the chosen parameterization. Updates within the parameterized set correspond to finding a stochastic policy with the smallest distance to the ground truth.}
  \label{fig:space}
\end{figure}

Policy Gradient (PG) Reinforcement Learning (RL) operates on the premise that it is possible to optimize a parametric policy by probing the environment, without the need to continuously update policy surrogate functions, and it constitutes an alternative to Q-learning algorithms \cite{sutton}. PGRL is naturally well suited to handle continuous actions in stochastic environments and, provided that the chosen policy is differentiable, the gradient of a cost function can always be estimated. The bias incurred by this method will also be related to the expressive power of the chosen policy (see Fig. \ref{fig:space}). RL has found multiple applications for quantum tasks, such as code design \cite{nautrup}, single qubit unitary compilation \cite{moro}, feedback control \cite{fosel,august}, and state preparation \cite{porotti}. Of particular interest are several works where the possibility of automatically learning how to optimize variational quantum algorithms \cite{artur, pgqaoa, yao, he} has been addressed. Interestingly, PG algorithms for variational quantum algorithms have been shown to be robust against noise. They can systematically outperform other methods in the training of small noisy circuits for combinatorial optimization. In this work, we explore a regime of shallow circuits with an increasing number of qubits, a regime complementary to that explored in \cite{pgqaoa}, where they focus on a QAOA ansatz for a small number of qubits.

In order to properly use RL in solving a quantum task, the agent's state, its available set of actions and the sampled reward need to be precisely defined (see Appendix A). In our case, the RL state will correspond to the quantum wavefunction, the set of possible actions $\theta = (\theta_1, \theta_2..., \theta_N)$ will be the set of free angles in $V(\theta)$, and the reward will be proportional to the state fidelity. It is useful to keep in mind that there is a conceptual gap between state and observation. Whereas the RL system is always in one state, the agent may or may not receive sufficient observations about that state. One could understand cast this learning task as a Quantum Partially Observable Markov Decision Process \cite{barry} in which (i) states are required to be pure, (ii) superoperators are required to be unitary, and (iii) except for the last step, the agent receives no rewards or observations about its environment. The simplest PGRL algorithm, known as REINFORCE \cite{williams}, draws extensively from Monte-Carlo learning, where episodes correspond to sampling at once all possible actions and then the performance is measured as the episode unravels \cite{sutton}.


\subsection{REINFORCE with Endpoint Rewards}

As motivated above, the REINFORCE algorithm can be slightly modified so that the reward is only related to the measured fidelity at the end of each circuit, meaning that we need not worry about performing intermediate measurements, which would collapse the wavefunction and interrupt the computation. The reward will be the overlap between the initial and the final state, given by Eq. (\ref{eq:fidelity}). During the first exploration of this approach, the set of possible actions $\theta = (\theta_1, \theta_2..., \theta_N)$ under the current policy, i.e. the angles of 1- and 2-qubit gates ($R_Y$ and $R_{ZZ}$ gates), will be randomly sampled from a Gaussian distribution of the form (see Fig. \ref{fig:setup}):

\be
\theta \sim \pi(x;\mu,\Sigma) = \frac{1}{\sqrt{2\uppi|\Sigma|}} e^{- (x-\mu)\Sigma^{-1}(x-\mu)^T},
\label{eq:gauss}
\ee
where the covariance matrix $\Sigma$ can either be fixed or obey some exploration-exploitation schedule, or it might even be learned(see Appendix A). The corresponding objective function is:

\bqa
J &=& \mathbb{E}_{\pi_\mu}[F] \nonumber \\
&=& \sum^m_k p_k \sum_\theta \pi(\theta|\mu,\Sigma) | \langle k|V(\theta)^\dagger U |k\rangle |^2,
\label{eq:cost}
\eqa
which corresponds to an average of the endpoint reward, i.e the asymptotic fidelity, over initial states (each sampled with probability $p_k$) and all possible actions (given by the current policy $\pi(\theta|\mu,\Sigma)$). In our case, maximizing $J$ corresponds to minimizing the associated cost function $J = \mathbb{E}_{\pi_\mu}[F]\stackrel[|\Sigma| \rightarrow 0]{}{\longrightarrow} 1 - C$.

The gradient of function in Eq.(\ref{eq:cost}) is:

\bqa
\nabla_\mu J &=& \sum^m_k p_k \sum_\theta \pi(\theta|\mu,\Sigma)   \nonumber \\
&&\times \nabla_\mu \log\pi(\theta|\mu,\Sigma) |\langle k|V(\theta)^\dagger U |k\rangle |^2,
\label{eq:gradient}
\eqa
where we have used the so-called ``policy gradient trick", which amounts to applying the chain rule to the policy function, and allows us to write the gradient of an expectation value as the expectation of a log-likelihood times a cost function \cite{sutton}. Estimating the gradient thus reduces to sampling a few episodes, i.e. performing several Monte-Carlo tree searches, and evaluating the cost function at the end of each one (see Appendix A).



%

\subsection{Random Walking over the Edge}

The cost landscape in many variational tasks is expected to become exponentially flat (in the number of qubits) except around some narrow gorges leading to good configurations. If, during optimization, the candidate solution finds itself in a non-zero slope region, sampling action configurations will very rapidly lead to non-vanishing gradients. One useful way of thinking about the Gaussian policy $\pi(\theta;\mu,\Sigma)$ is as quantifying the probability of obtaining a configuration that is at some distance from a minimum at location $\mu$. With this parameterization, the RL problem will learn to place $\mu$ as close as possible to the nearest minimum in the current non-zero slope region.

If, on the contrary, the current $\mu$ is in a flat region, which is overwhelmingly more likely to happen, it will be very far away in Euclidean distance from any minimum. In this case we do not expect our method to perform better than average as the dimensionality grows. In this case the gradient will consist of a (constant) cost times an empirical expectation of a Gaussian displacement:

\bqa
\nabla_\mu J&=& \sum_\theta  \sum^m_k p_k \pi(\theta|\mu,\Sigma) \nonumber \\
&&\times \nabla_\mu \log\pi(\theta|\mu,\Sigma)  |\langle k|V(\theta)^\dagger U |k\rangle |^2\nonumber \\
&\approx& \frac{1}{N_{eps}}\sum^{N_{eps}}_\theta  [\Sigma^{-1}(\theta - \mu)] \times \epsilon, 
\label{eq:gradient2}
\eqa
where the constant $\epsilon$ corresponds to the fidelity evaluated in the flat region. This will lead to a random walk for which one can compute the mean square displacement to be $\sim\eta\epsilon \sqrt{\frac{N_{iters}}{N_{eps}} \Tr \Sigma^{-1}} $, with $\eta$ the learning rate of the gradient descent algorithm (see Appendices A and C) and $1/N_{iters}$ the discretized time lapse. As explained in the Appendices, the counter-intuitive presence of an inverse covariance in the mean square displacement is due to the fact that the update rule is inversely proportional to the policy's current covariance, as it would otherwise favour configurations that are frequently sampled, rather than those with high rewards \cite{sutton}. We have numerically verified that, for initializations within flat cost landscapes, the performance of PG gets degraded as the eigenvalues of the covariance matrix $\Sigma$ grow, in accordance with the expression for the mean-squared displacement in Appendix C. This random walk evolves within a hyperball with radius increasing roughly as $\sim \sqrt{N_q D N_{iters}}$, with $D$ the circuit depth. However, since the volume ratio of a hyperball and its corresponding hypercube vanishes with a factorial dependence on the dimensionality, this approach is expected to stall deep inside a barren plateau. 

\begin{figure}[ht!]
  \includegraphics[width=\linewidth]{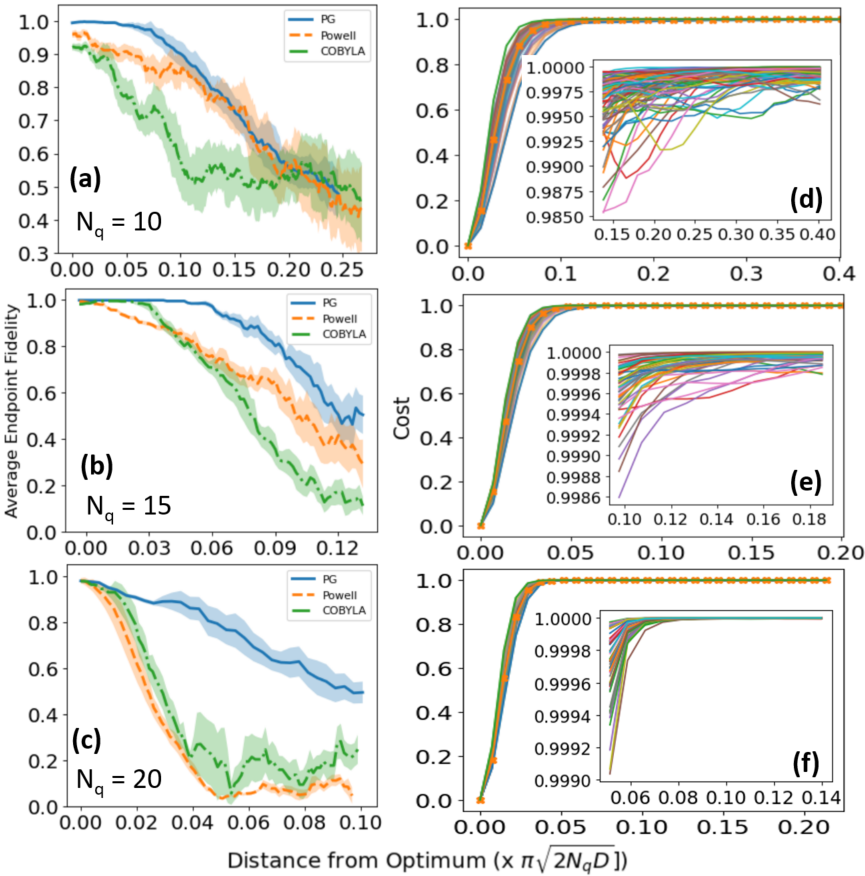}
  \caption{Performance, measured as the asymptotic reward, as a function of the initialization distance $d = |\hat \theta - \theta^*| / (\pi\sqrt{2N_q D})$ over the torus. Several circuits were run for increasing distances and moving averages taken over the whole sequence of {\bf (a)} 200 runs for $N_q = 10$ and $D=3$, {\bf (b)} 100 runs for $N_q = 15$ and $D=4$, and {\bf (c)} 50 runs for $N_q = 20$ and $D=5$. The shadows correspond to one moving standard deviation. PG with endpoint rewards typically performs better than COBYLA and Powell optimisers. Our numerics show that the improvement in performance becomes more dramatic as the number of qubits grows. This can be seen as evidence that whereas gradient-free methods are unable to move in the correct direction near the onset of the barren plateau, PG updates in the diffusive regime allow it to explore increasingly far configurations and eventually find a non-zero slope region (see Fig.\ref{fig:BPtrajectories}). {\bf{(d-f)}} Numerically computed cross-sections along random directions of the cost landscape. The gorge corresponding to the global optimum gets narrower as the number of qubit increases. Insets: the exponential reduction of the cost variance. Whereas for {\bf{(d)}} 10 qubits the fluctuations in the cost remain in the order of $1\%$, which can be fully exploited by gradient-free optimizers, for {\bf{(f)}} 20 qubits the cost fluctuations rapidly descend well below $10^{-4}$.}
  \label{fig:performance}
\end{figure}

The regions of interest are in the cross-over between the two regimes. In those regions (i.e. at the edge of, yet within a flat landscape)  PG-based training can ``feel" a change in slope if allowed to diffuse for sufficiently long. In Fig.\ref{fig:performance} we illustrate the region where enhanced performance is expected, where the candidate solution random-walks over the edge and enters a non-zero slope region.

We argue that the correct ensemble of benchmarks for this method are gradient-free optimizers, as opposed to gradient-based methods (based for instance on the parameter-shift rule \cite{schuld}). Contrary to what happens in classical neural networks (where back-propagation can be used), the gradient calculated in VQAs needs to be done using several of the equivalent of forward-backward passes, which amounts to computing commutators with some real or fiducial Hamiltonian (see for example \cite{sharma} ). This means that, for each iteration, one commutator per free parameter has to be evaluated. Moreover, this evaluation needs to be done within a fixed tolerance. So the number of measurements per commutator grows as the inverse tolerance squared times the number of free parameters.
In addition, PG optimization of variational algorithms is a non-local optimization procedure, since estimating the gradient of the cost function involves sampling episodes in the vicinity of the current tentative solution. This is aligned in philosophy with gradient-free optimizers which sample finite differences relative to different locations in parameter space, rather than gradient descent, in which the gradient is evaluated at a single point in parameter space for each iteration.

Our aim will be to empirically compare the performance of PG-based variational compilation with gradient-free methods.

\section{Numerical Experiments and Results}

To assess the performance of PG methods applied to variational compilation, we have run numerical simulations of the training procedure, both in the noiseless case and for noisy circuits. We generated several random shallow quantum circuits with depths logarithmic in the number of qubits and known connectivity graphs, which acted as the target unitary $U$, followed by a circuit with the same connectivity graph and depth, and randomized parameters implementing another unitary $V$. As hinted previously, this setup is physically motivated because in the absence of error correction, the circuit depth of NISQ algorithms is bounded by the inverse effective noise rate. This means that only shallow circuits, i.e. of constant depth, can be realistically considered \cite{bravyi18, bravyi20}. Moreover, the logarithmic depth regime is expected to suffer from the barren plateau effect for global cost functions \cite{cerezo}.

Practically, our choice of this setup stems from the need to evaluate how close the performance gets to its theoretical maximum. Given that most unitaries have exponentially long circuits \cite{NCbook}, sampling operators in $SU(2^{N_q})$, instead of explicitly defining a quantum circuit, would almost certainly result in the optimization getting stuck at indeterminate values of the cost function, which would in turn lead to poor characterization of the performance.

Our methodology consisted of testing different gradient-free methods, such as Powell (which relies on bi-directional search) and COBYLA (simplex-based), and a simple variation of REINFORCE PG, and to compare their relative performance in training $V(\theta)$ to emulate the inverse of the target unitary, i.e to determine $\theta^*$ such that $V(\theta^*) = U^\dagger$. We used the {\em Cirq} simulator for this purpose \cite{cirq}.

In order to establish a meaningful comparison between PG-based training and gradient-free optimisers, it is necessary to quantify the resources that either method needs to converge. Each iteration of a gradient-free optimiser entails a fixed number of runs ($n_{shots}$) of a quantum processor. In PG-based training, each episode involves sampling $N_{eps}$ configurations in the vicinity of the current configuration to estimate the gradient, so the number of runs is $n_{shots} \times N_{eps}$. Two factors render the comparison difficult. The first one is that the learning rate is a hyperparameter that can be tuned, and the number of iterations depends heavily on it. The second consideration is that PG-based training is robust to fluctuations (see Appendix D), so its performance is not degraded as much as that of gradient-free optimisers as $n_{shots}$ is reduced (see  Fig. \ref{fig:noise}). As a general rule, we have set a maximum number of iterations to a sufficiently large value so that COBYLA and Powell typically converge (either to good or bad solutions). We found empirically this to grow very fast on the number of qubits, from about a few hundred iterations for 10 qubits to about $\sim 10^4$ for 20 qubits.

\subsection{Performance vs Distance from Optimum}

\begin{figure}[ht!]
  \includegraphics[width=\linewidth]{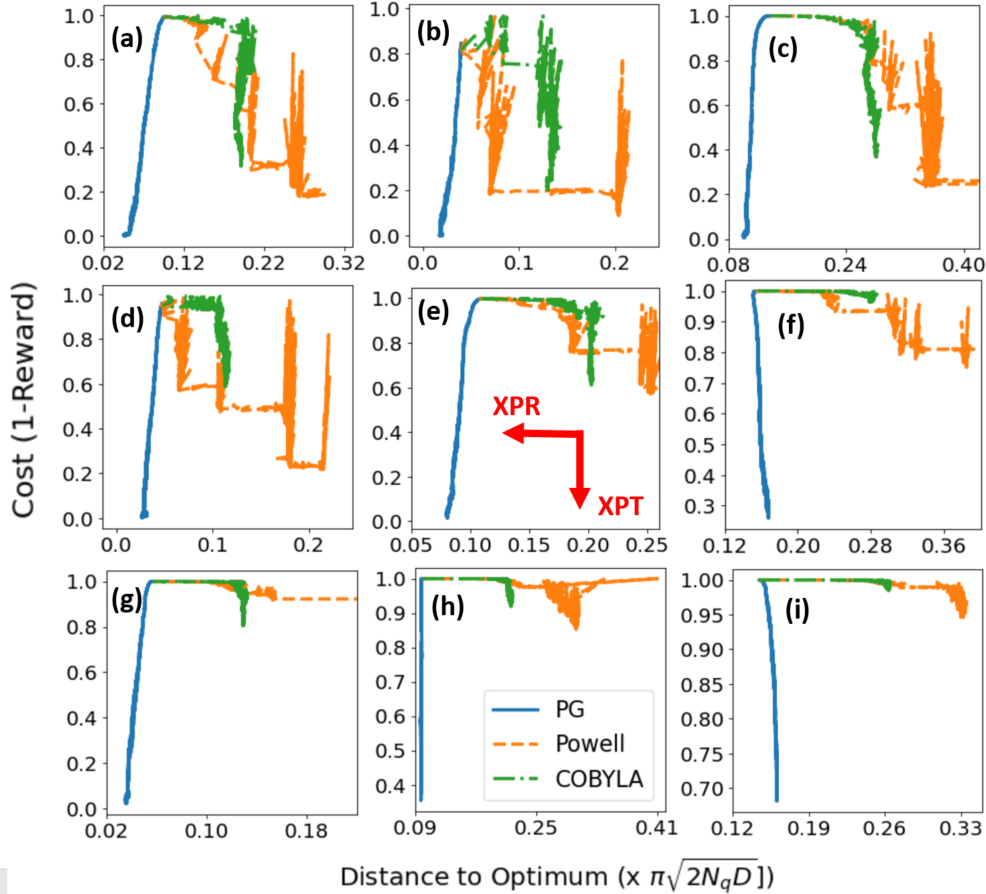}
  \caption{Optimization trajectories, depicted in $(distance, cost)$-coordinates for 10 {\bf{(a-c)}}, 15 {\bf{(d-f)}} and 20 qubits {\bf{(g-i)}} and approximate initialization distances $0.05$ {\bf(a,d,g)}, $0.1$ {\bf(b,e,h)} and $0.15$ {\bf(c,f,i)}. In these coordinates, it is possible to diagnose whether the optimization is in an exploration (XPR) or an exploitation (XPT) phase. The exploration phase is characterized by searching for new configurations even if it does not result in a net cost reduction. Conversely, in an exploitation phase, priority is given to updates that minimise the cost (see {\bf (e)}). For a fixed number of 5000 runs per iteration, COBYLA and Powell optimisers are not able to ``feel" the slope and rapidly get stuck in local minima. In subplots {\bf(f, h, i)} the PG optimizer has clearly been trapped in a local minimum, as can be gleaned from the slope of the optimization trajectory. Whereas COBYLA, based on the simplex method, features a zig-zag behaviour typical of pivot operations, each iteration of the Powell method involves a line-wise minimization. Both of these optimizations are discrete in the sense that each update can bring the current configuration to a very different position in $\theta$-space. During PG-based learning, the candidate configuration is updated ``continuously" if the learning rate is sufficiently small. This, together with the fact that, by tuning the covariance configurations beyond local maxima can be sampled, allows for a smoother optimization trajectory, resulting in a more balanced alternation between exploration and exploitation.}
  \label{fig:BPtrajectories}
\end{figure}

The trainability of variational circuits, measured in terms of the obtained reward, as a function of the distance from the optimum, is shown in Fig.\ref{fig:performance}. The performance of policy gradient is degraded further away from the optimum, albeit significantly less than for gradient-free optimizers, as the tentative solution is initialized further and further from the optimum. We interpret this as a consequence of the PG algorithm entering a ``diffusive regime" in which the tentative solution performs a random walk and at some point falls over the edge.  Evidence for this behaviour is provided by the trajectories shown in Fig.\ref{fig:BPtrajectories}. There, one can see that, whereas simplex-based methods are unable to move in the correct directions if initialized too far from the optimum (cf. \cite{arrasmith}), the PG-based optimization is able to perform a more accurate exploration of the parameter space.

Gradient-based optimizers will also perform a random walk in a barren plateau if the magnitude of the gradient is too small compared to the measurement precision \cite{skolik}, however this behaviour only depends on evaluations of the gradient at a single point in parameter space. On the contrary, the random walk performed by the REINFORCE algorithm will explore configurations in a non-local way (by sampling episodes at in the vicinity of the current configuration), whose performances are then averaged and used to perform an update.

\subsection{Training Noisy Circuits}

Noise can be shown to impose an upper bound on the gradient strength and it further hinders the trainability of variational quantum circuits of linear depth \cite{wang}. We have sought to estimate the robustness of the method in the presence of noise in the logarithmic depth regime. The noise simulations involved appending a single-qubit depolarizing channel after each single qubit gate. After each depolarizing channel, the average loss of fidelity is $(1-p)$, so after D layers, the fidelity will decrease on average by $(1-p)^{N_q D}$, irrespective of the angle.  Each of the episodes needed to estimate the gradient will sample a reward that will be reduced by the same factor on average. Being a stochastic update policy, PG is naturally robust against this kind of errors, so the search will proceed in the correct direction, albeit with more variance and slower convergence (see Appendix D). Fig.\ref{fig:noise} depicts the relative performances of PG-based optimization, COBYLA, and Powell in the presence of noise.

\begin{figure}[ht!]
  \includegraphics[width=\linewidth]{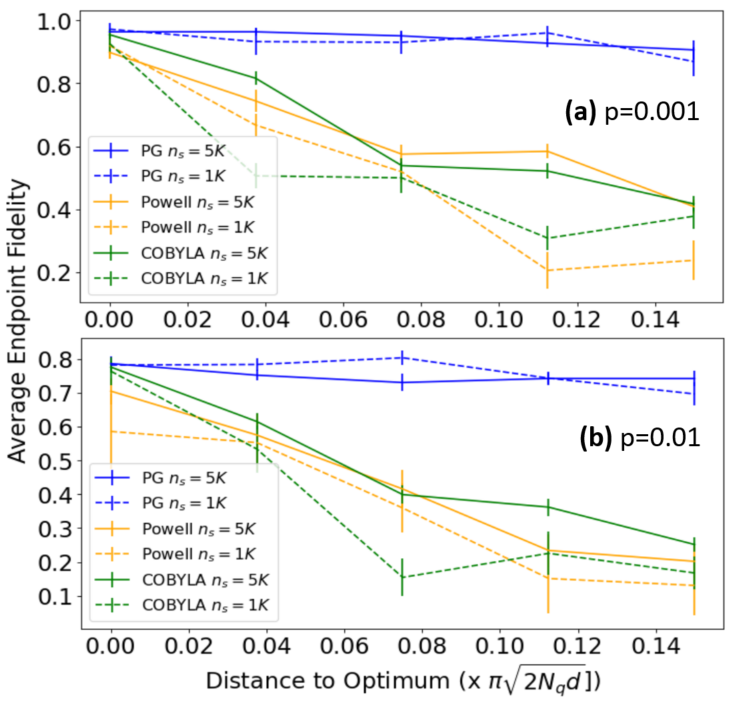}
  \caption{Training of circuits of 10 qubits in the presence of depolarizing noise with probabilities {\bf (a)} $p=0.001$ and {\bf (a)} $p=0.01$ after each single qubit rotation. Each optimization involved either $n_{shots} = 5000$ or $n_{shots} = 1000$ per iteration. Error bars denote averaging over 10 optimizations. Compared to performances in Fig. \ref{fig:performance}, one can see that PG is more robust than COBYLA and Powell methods in landscapes of non-zero slope. PG averages noise in gradients over successive updates, provided that noise is independent of $\theta$ (see Appendix D). The damping of the endpoint fidelities is proportional to $(1-p)^{N_q D}$. Increasing the number of shots can reduce the fluctuations at each step, leading to better convergence to the noisy maximum.}
  \label{fig:noise}
\end{figure}

The observed behaviour in the numerics is again consistent with PG-based optimization being able to find good solutions ``at the edge" of a barren plateau even in the presence of noise. This can be understood as a consequence of the optimisation taking steps in the correct direction on average. If the step size ($\eta$) is sufficiently small, fluctuations will average out over successive steps. 

\subsection{Generalization Error}

In order to estimate the reward (circuit endpoint fidelity) without an ancillary register and long range interactions, which are needed to implement the Hilbert-Schmidt test \cite{sharma, wang}, we trained the circuit on small subsets of initial states and estimated the fidelity (see Appendix B). It is important to assess the generalization error of this method on states not included in the training of the variational circuit.

\begin{figure}[ht!]
  \includegraphics[width=\linewidth]{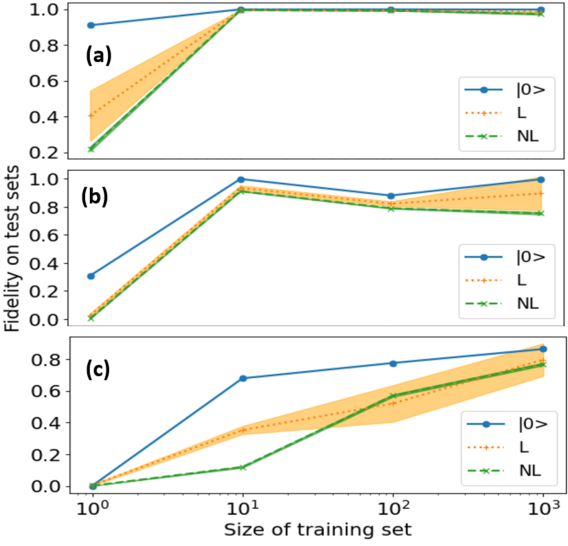}
  \caption{Generalization error on different test sets, for circuit configurations initialised at increasing distances {\bf (a)} 0.05, {\bf (b)} 0.1 and {\bf (c)} 0.15  $(\times \pi\sqrt{2N_q D})$ from the known optimum. Shading corresponds to one standard deviation. Training used $n_{shots} = 2000$ circuit runs per rollout. Larger training sets demanded more control of the learning rate to improve, and the number of iterations ranged from a few hundred (for a training set with only one initial state) to a maximum of 5000 iterations (for 1000 initial states). The local test set was made of 100 states of the form $\bigotimes^{10}_j \ket{\psi}_j $ where $\ket{\psi}_j$s are random qubit states, whereas the non-local test set was made of 100 random state vectors}
  \label{fig:generalization}
\end{figure}

We trained several 10 qubit circuits on increasingly large sets of initial states. Their performance was then tested on different test sets and distances from the optimum. The training set was made of local quantum states of the form $\ket{k} = \bigotimes^{10}_j R_Y(\frac{\pi}{4}\times c_j)\ket{0}_j $, where $c_j$ are random integers in the range $[0,2^3 -1]$. The circuit was tested on the $\ket{0}$ state, which was not necessarily included in the training set, and two additional sets: one made of tensor products of local rotations and another one made of random state vectors (which are non-local with high probability). As shown in Fig. \ref{fig:generalization}, the performance on test sets increases as the number of states in the training set grows, in accordance with calculations in Appendix B.

\section{Conclusion}

We have introduced a method to compiling variational quantum circuits using techniques from reinforcement learning. This approach constitutes an alternative to gradient-free methods and has the potential to outperform them on the edge of a barren plateau. This is because the optimization is not performed over $\theta$-space, but rather in the space of statistical distributions over $\theta$, for which it is possible to implement classical gradient ascent methods for the distribution parameters ($\mu$ in our Gaussian case). A differentiable expression for the cost allows us to estimate the gradient via non-local evaluations of the cost function. 

Another salient feature of PG for variational compilation is that its performance is robust to noise. This crucial property allows us to achieve better performances in simulations with depolarizing noise strengths that are commensurate with state-of-the-art gate fidelities \cite{superconducting, silicon1,silicon2, trappedions}.

This method has the potential to be used as an efficient way to train shallow quantum circuits (for up to about a few tens of qubits) in the presence of noise, which could in turn be used as building blocks for larger circuits. An intriguing question is whether this method could be used as the lowest level of recursion in compilers reliant on the Solovay-Kitaev construction, where it could be used to replace the need for a library of gate sequences (along the lines of \cite{moro}). Improvements to the method introduced in this work could combine temporal difference learning with policy gradient, such as actor-critic methods, which could be used to train circuits layerwise (similar to the methods in \cite{artur} and \cite {skolik}).

While it is difficult to compare the runtimes of different approaches, we found that, for a fixed performance threshold, PG-based approximate compilation is typically more efficient both in terms of absolute time and number of queries to a quantum computer, than the gradient-free methods we considered, since the episode sampling can easily be parallelized. This is the case for larger circuits and in the presence of depolarizing noise.

Finally, we expect this RL-based approach to circuit training to be beneficial in other quantum variational tasks in addition to unitary gate compilation. Quantum circuits with a number of parameters that grows slowly in the circuit depth, such as QAOA, are naturally better suited to this method than quantum tasks in which the number of parameters is linear or polynomial in the depth.

\section{Code Availability}

Part of the code used to generate these results is available upon reasonable request.




\onecolumn{
\bibliographystyle{unsrt}

}

\widetext


\section*{Appendix A: \\
Policy Gradient with Gaussian Policies and RMSprop}

The first step towards phrasing our approximate compilation problem as a reinforcement learning problem is to map the process of training a variational quantum circuit to a Markov Decision Process (MDP). Different mappings will lead to different learning scenarios in which distinct facets or challenges of the original quantum task will become apparent. An MDP consists of a tuple $(\mathcal{S}, \mathcal{A},  R_{s,a}, P_{s',(s,a)},)$, with $\mathcal{S}$ the set of states, $\mathcal{A}$ the set of actions, $R$ is the reward obtained by taking action $a$ in state $s$, and finally $P$ is a stochastic matrix giving the probability of transitioning to state $s'$ given that the current state is $s$ and the current action is $a$. Generally, P is so large that it can only be sampled by an agent exploring an environment. An agent seeking to maximize the long-term reward of an MDP can do so by optimizing a policy $\pi(a,s)$, which associates a probability to each available action-state pair.

There exist several approaches to optimizing a policy. Temporal Difference methods aim at measuring the reward after each transition and update the value of each state under the current policy (the estimated long-term reward associated to that state). Optimality of the corresponding policies is ensured by the Bellmann Optimality Condition \cite{sutton}. Another approach is given by the direct optimization of the policy, thus relying little, or not at all, on value iteration. One of the simplest policy gradient algorithms, conceptually as well as in terms of implementation, is the REINFORCE algorithm \cite{williams}. In REINFORCE, the policy is parameterized and belongs to a variational family of distributions, such that it is possible to differentiate it with respect to the variational parameters. 

\subsection*{Reinforce with Endpoint Rewards}

Throughout this work we have used a variation of the REINFORCE algorithm in which the reward will only be obtained at the end of a given sequence of actions (an episode). The motivation for this choice is ease of comparison with gradient-free optimisers, as explained in the main text. The rewards gathered at the end of each episode are used to estimate the gradient as follows: at each iteration, the agent performs a Monte-Carlo tree search, i.e. it explores the space of actions for a fixed amount of time.

\be
J = \mathbb{E}_{\pi_{\mu,\Sigma}}[R] = \sum_\theta \sum^m_k p_k \pi(\theta|\mu,\Sigma)  r^{(k)}_\theta,
\label{eq:reward}
\ee
where we start at state $\ket{k}$ with probability $p_k$ and the reward $r^{(k)}_\theta = |\langle k|V(\theta)^\dagger U |k\rangle|^2$ is related to the end-state fidelity. Now the problem is to compute the gradient of this expectation value. Each particular trajectory in the Monte-Carlo tree search, a \emph{rollout} in the RL jargon, contributes with a weight that is proportional to its associated final reward. This can be seen by applying the chain rule to the policy function:

\be
\nabla_\xi J =\sum_\theta \sum^m_k p_k \pi(\theta|\mu,\Sigma) \nabla_\xi \log\pi(\theta|\mu,\Sigma)   r^{(k)}_\theta,
\label{eq:grads}
\ee
where $\xi\in\{\mu,\Sigma\}$ and we have used the Gaussian policy:

\be
\pi(x;\mu,\Sigma) = \frac{1}{\sqrt{2\uppi |\Sigma|}} e^{- (x-\mu)\Sigma^{-1}(x-\mu)^T}.
\label{eq:policy}
\ee

The ``Log-likelihood trick" allows us to express a gradient of an expectation value as the expectation value of a different gradient, which can be estimated numerically. The logarithm of the Gaussian policy has the following gradients:

\bqa
\nabla_\mu\log\pi(x;\mu,\Sigma) &=& \Sigma^{-1}(x-\mu),  \\
\nabla_\Sigma \log\pi(x;\mu,\Sigma) &=&  -\frac{1}{2} \Sigma^{-1} (\mathbf{1} - (x-\mu)^T(x-\mu)\Sigma^{-1}),
\label{eq:gradMuSigma}
\eqa
which allows for learning both the mean $\mu$ and the covariance $\Sigma$. However, in this work we do not learn $\Sigma$ but rather fix a simple exploration-exploitation schedule $\Sigma(t) = (1-t/T) \Sigma_i + t/T \Sigma_f$, such that $\Sigma_i \gg \Sigma_f\longrightarrow 0$.

\subsection*{Reinforce with Intermediate Rewards}

One possible improvement over REINFORCE with endpoint rewards is to perform intermediate measurements of the quantum state. The overhead incurred by this method is proportional to the number of layers, so in our case, we argue that this is a reasonable cost. If we do perform projective measurements onto the wavefunction after each layer $V^{(layer s)}(\vec\theta_s)$ of the $V(\vec\theta_T) = V^{(layer D)}(\vec\theta_3) \circ  \dots V^{(layer 2)}(\vec\theta_2) \circ V^{(layer 1)}(\vec\theta_1) $ circuit, such that $\vec\theta_T = [\vec\theta_1,\vec\theta_2...,\vec\theta_D]$, then the generalised cost function is:

\be
J_{LW} = \mathbb{E}_{\pi_{\mu,\Sigma}}[R] = \sum^m_k p_k \sum^D_s  \sum_{\theta_s}\pi(\theta_s|\mu_s,\Sigma_s)  G^{(k)}_s,
\label{eq:reward_episodic}
\ee
where LW stands for layerwise and the index s denotes the layer after which the fidelity is evaluated, ranging from 1 to a maximum depth of $D\propto \log N_q$ and the generalized return is $G^{(k)}_s = \frac{1}{D-s+1}\sum^D_{s'>s} \gamma^{s'-s}|\langle k|V(\vec\theta_{s'})^\dagger\circ\dots\circ V(\vec\theta_1)^\dagger U |k\rangle |^2$.  The gradient is taken layerwise:

\be
\nabla_{\mu_s} J_{LW}= \mathbb{E}_{\pi_{\mu_s,\Sigma_s}}[R] = \sum^m_k p_k  \sum^D_s \sum_{\theta_s} \pi(\theta_s|\mu_s,\Sigma_s)\nabla_{\mu_s} \log\pi(\theta_s|\mu_s,\Sigma_s)  G^{(k)}_s.
\label{eq:gradient_episodic}
\ee

Setting $\gamma=1$ corresponds to a far-sighted policy, which incentivizes the $V(\vec\theta_T)$ circuit to ``unscramble" the computation as much as possible at each step while keeping a concerted global action. This strategy is similar in spirit to that followed in \cite{artur}, in which the state of the RL agent corresponds to the wavefunction and has to be measured after each set of actions to sample the reward.  Measuring the reward after each layer, for logarithmic depth circuits adds a relatively low overhead and it has the potential to lead to higher performances, especially if supplemented with temporal-difference techniques.

\subsection*{RMSprop and Baseline}
 
Once the gradient has been estimated, updates can be performed following a \emph{RMSprop} gradient update rule \cite{goodfellow}. RMSprop is an adaptive learning rate method with empirically better convergence properties than simple gradient ascent methods in deep learning. It works by (i) computing a discounted moving average of the gradient variances and (ii) dividing the update step by the discounted variance. The result is that the learning rate will increase in relatively flat landscape directions and it will decrease in steep directions. Let  $\sigma^{2}$ be the variance of the computed gradients over different episodes. 

\bqa
\sigma^{2, (t)}_j &=& \upgamma  \sigma^{2, (t-1)}_j + (1 - \upgamma) (\nabla_\xi J^{(t)})_j^2,\\
\xi_j &\leftarrow& \xi_j + \eta \frac{(\nabla_\xi J^{(t)})_j}{\sqrt{\sigma^{2, (t)}_j + \varepsilon}},
\eqa
where $\upgamma$ is the discount factor. This update rule has been empirically shown to allow for a more efficient exploration of complex cost landscapes. A table with the parameters used for this work is provided below:

\begin{center}
\begin{tabular}{||c | c ||} 
 \hline
 Parameter & Value \\ [0.5ex] 
 \hline\hline
 $\Sigma_i$ &  diag$(5\times 10^{-3})$  \\
 \hline
 $\Sigma_f$ &  diag$(10^{-6})$  \\
 \hline
 $N_{eps}$ & 5-10  \\
 \hline
 $p_s$ & 1/m  \\
 \hline
 m & $ N_q^2 $ \\ [1ex] 
 \hline
$\upgamma$ &  0.9 \\ 
 \hline
 $\eta$ &  $5\times10^{-3}$ - $1\times10^{-4}$ \\
 \hline
 $\varepsilon$ & $10^{-8}$  \\
 \hline
$maxiter_{\textrm{REINFORCE}}$ & $1000-10000$  \\
 \hline
$maxiter_{\textrm{COBYLA}}$ & $10000$  \\
 \hline
$maxiter_{\textrm{Powell}}$ & $30000$  \\
 \hline
\end{tabular}
\end{center}

A further consideration is the choice of the baseline $b$ in the computation of the gradient, which allows us to reduce the variance of the estimator. Provided the baseline is positively correlated with the end reward, i.e. if $Cov[R, b]>0$, the advantage $(R-b)$ will have less variance as the baseline will compensate for fluctuations of $R$. We have used a simple mean, i.e. $b=\frac{1}{m} \Sigma_\theta r_\theta$, which intuitively gives an advantage for the rollouts which perform better than average. Reducing the training error involves tuning all the hyperparameters of the policy gradient algorithm, for which there are several prescriptions to follow, see \cite{bayesianopt} for details.

\section*{Appendix B: \\
Efficient Trainability without Tomographically Complete Measurements}

It is in principle possible to efficiently calculate the fidelity at the end of each Monte-Carlo rollout by means of the so-called Hilbert-Schmidt test \cite{khatri, sharma}, which makes use of an array of ancilla qubits and a series of potentially long range interactions.

In the absence of a secondary qubit register, one needs to resort to an estimation of the fidelity via Eq.(\ref{eq:fidelity}). In the noiseless case, it is possible to recover the exact averaged value of the fidelity $F(\theta) = \frac{1}{4^{N_q}} \sum^{\sim {N_q}}_k|\langle k|V(\theta)^\dagger U |k\rangle |^2$ using a tomographically complete characterization.

An important consideration, therefore, is to ensure that our estimate of the fidelity $\hat F(\theta)$, which makes use of a massively downsampled subensemble of orthogonal states, is sufficient to train our variational circuit.

The set of initial states, $\{\ket{k}\}^m$ with $\braket {k_i}{k_j}=\delta_{ij}$, allows us to build an estimator $\hat F = \frac{1}{m}\sum^m_k f_i$, where $0\leq f_i\leq 1$ can be interpreted as independent samples of the true fidelity. In the limit where the fluctuations due to finite sampling vanish (infinitely many repetitions for each Monte-Carlo rollout), the Hoeffding inequality implies:

\be
P(|\hat F - F| \geq \epsilon) \leq 2 \exp - 2\epsilon^2 m.
\ee

For large numbers of qubits, small differences in fidelity become increasingly significative and difficult to obtain through numerical optimization, which amounts to imposing a scaling on the estimator accuracy,  $\epsilon \sim N_	q^{-t}$ for some positive real $t$. To satisfy the regime $ \epsilon^2m\gg 1$ for any number of qubits gives a scaling of $m\sim\Omega(N_q	^{2t})$.

\section*{Appendix C: \\
N-dimensional Random Walk}

We will now present some basic facts about random walks in high-dimensional spaces, as this is the regime that describes a stalled optimization deep inside a barren plateau. Within an exponentially flat landscape and under a Gaussian policy, the estimator of the gradient can be written as:

\be
\hat{\nabla_\mu J} = \frac{1}{N_{eps}}\sum^{N_{eps}}_\theta  [\Sigma^{-1}(\theta - \mu)] \times \epsilon,
\ee
where $\epsilon$ corresponds to the reward (fidelity). The expected displacement at each iteration will be $\delta \mu = \eta \hat{\nabla_\mu J}$, whose covariance can be calculated to be $Cov[\delta \mu] = \eta^2 \epsilon^2 \Sigma^{-2}\hat\Sigma \approx \eta^2 \epsilon^2 \Sigma^{-1} $. It is at first surprising that the variance of the expected displacement is inversely proportional to the initial covariance matrix. This is due to the fact that, in a Gaussian policy, the update rule is proportional to the reward but inversely proportional to the variance of exploration. Were this not the case, angles that are selected more frequently (as directions in $\theta$-space with larger variances will be sampled more frequently) might push the learning in a direction which is not that of the highest returns. This can be traced back to the relation between the logarithm of the parameterized Gaussian and the inverse probability of an action under a fixed policy\cite{sutton}.

For independent updates, i.e. $\langle \delta\mu_{(i)}\delta\mu_{(j)}\rangle = \delta\mu_{(j)}^2\delta_{ij}$, at is the case for REINFORCE in an exponentially flat region, one can express the mean square displacement ($\bar{\delta\mu}^2$) of the resulting random walk as:

\bqa
\bar{\delta\mu}^2 &=& \langle(\sum^{N_{iters}}_j\delta\mu_{(j)})^2\rangle = \langle\sum^{N_{iters}}_j\delta\mu_{(j)}^2\rangle = \sum^{N_{iters}}_j \langle\delta\mu_{(j)}^2\rangle \nonumber \\
&=& N_{iters} \frac{\eta^2\epsilon^2}{N^2_{eps}} \langle[\sum_{ij}\Tr \Sigma^{-2}(\theta_i-\mu)(\theta_j-\mu)^T \rangle \nonumber\\
&\approx& N_{iters} \frac{\eta^2\epsilon^2}{N_{eps}} \Tr \Sigma^{-1} \leq N_{iters} \frac{\eta^2\epsilon^2}{N_{eps}} \frac{2N_q D}{\sigma^2_{MIN}},
\eqa
where the angle brackets denote an ensemble average, and we have used that the estimator of the covariance matrix is $\hat \Sigma = \frac{1}{N_{eps}}\sum_{ij}(\theta_i-\mu)(\theta_j-\mu)^T\stackrel[N_{eps}>>1]{}{\approx}\Sigma$. The number of free parameters is $n_{params} = 2N_q D$ and $\sigma^2_{MIN}$ is the minimum variance across all dimensions within the policy. The fact that the inverse of the covariance appears in the mean square displacement calculation might seem counterintuitive, as one would expect a diffusive process to be proportional to the strength of the fluctuations rather than to their inverse. However, as explained above, the update rule given by the gradient estimator involves weighting by the inverse of the covariance to compensate for actions that are too likely to happen under the current policy.

This diffusive character of the optimization can be interpreted as a  local search in an $n_{params}$-dimensional neighborhood performed by a random walker. The radius of the explored hyperball grows with the square root of the number of iterations. However, in high-dimensional spaces, the volume ratio between a hyperball of typical dimension $\propto\sqrt{N_{iters}}$ and the corresponding hypercube vanishes as $\pi^{n_{params}/2} / \Gamma(\frac{n_{params}}{2} + 1)$, with $\Gamma(x)$  the Euler's gamma, which is the reason why local search stalls deep inside in a barren plateau.

\section*{Appendix D: \\
Error Propagation for Depolarizing Channel}

In our error model, the depolarizing channel $\hat C_p\rho = (1-p)\rho + p I/2$  acts on each qubit after every single-qubit gate, giving rise to an effective noise rate $p_{eff} = 1 - (1-p)^{N_q D}$, with $N_q$ the number of qubits and $D\sim \log N_q$ the depth of the circuit. The fidelity measured for a single episode in each PG iteration is expected to be damped according to: 

\be
F_{noisy} \propto (1 - p_{eff}) F_{noiseless}.
\ee

This simple phenomenological model gives rise to a gradient update rule that is unbiased with respect to the noiseless one, with deviations that decrease proportionally to the square root of the number of shots (circuit repetitions).

The ratio $\alpha = \delta \mu_{noisy}\delta \mu_{noiseless} / |\delta \mu_{noisy}||\delta \mu_{noiseless}|$ can be used to compare the noisy update rule to the noiseless one. This quantity can be understood as a cosine similarity measure, such that whenever the vectors are parallel (or anti-parallel), it is maximal, and it is zero if the vectors are orthogonal. Considering $\hat p_{(\theta),g} = n^{(\theta),g}_{errors}/n_{shots}$, where $n^{(\theta), g}_{errors}$ corresponds to the number of depolarizing errors in a particular instantiation of a circuit gate, and, furthermore, considering that for Gaussian policies one can express an update as $\delta \mu_{noiseless} = \eta / N_{eps} \sum_i  [\sum_\theta F_\theta  \Sigma^{-1}(\mu - \theta)]_i$ where $[...]_i$ is an unnormed coordinate vector and $F_\theta$ is the fidelity corresponding to a given circuit/episode: 

\bqa
\mathbb{E} [\alpha] &=& \mathbb{E}\frac{\eta^2/N^2_{eps}\sum_i  [\sum_\theta (1-\Pi^{N_q D}_g(1-\hat p_{(\theta),g})) F_\theta  \Sigma^{-1}(\mu - \theta)]_i \cdot [\sum_\theta F_\theta  \Sigma^{-1}(\mu - \theta)]_i}{|\delta \mu_{noisy}||\delta \mu_{noiseless}|} \nonumber \\
&=& \frac{\eta^2/N^2_{eps}(1-(1- p)^{N_q D})\sum_i  [\sum_\theta F_\theta  \Sigma^{-1}(\mu - \theta)]^2_i }{(1-(1- p)^{N_q D})|\delta \mu_{noiseless}|^2} \\
&=& 1\nonumber ,
\eqa
since we are assuming $\mathbb{E} [\hat p_{(\theta),g}] = p$ for all possible angle configurations and all qubits. This means that on average, the noisy update points in the correct direction. This is because the fidelity is assumed to be damped by the same ratio for all angles. Error propagation gives:

\be
\Delta F_{noisy} = |\partial F_{noisy} / \partial p| \Delta p \propto N_q D p (1-p) \frac{\sigma_{F}}{\sqrt{n_{shots}}},
\ee
with $\sigma_{F}$ arising from quantum noise. Increasing the number of repetitions will therefore narrow down the variance at each step in the gradient update.

This reasoning is not expected to hold for noise models where $\mathbb{E}[\hat p_{(\theta),g}]$ is not independent of the angles (for instance for the low-temperature amplitude damping channel).

\end{document}